\documentclass{cs19proc}

\newcommand{\Teff}{\ensuremath{T_{\rm eff}}}                      
\newcommand{\er}[3]{\ensuremath{#1^{+#2}_{-#3}}}                  
\newcommand{\Prot}{\ensuremath{P_{\rm rot}}}                      
\newcommand{\Vsini}{\ensuremath{v \sin i}}                        
\newcommand{\degr}{\ensuremath{^{\circ}}}                         
\newcommand{\Mjup}{\,\ensuremath{\rm M_{\rm Jup}}}                
\newcommand{\Rjup}{\,\ensuremath{\rm R_{\rm Jup}}}                
\newcommand{\Porb}{\ensuremath{P_{\rm orb}}}                      
\def\fo{\hbox{$\,.\!\!\!^\circ$}}                                 
\def\fd{\hbox{$\,.\!\!\!^{\rm d}$}}                               

\editors{G.~A. Feiden}
\publisher{Zenodo}
\conference{The 19th Cambridge Workshop on Cool Stars, Stellar Systems, and the Sun}
\conferencedate{2016}

\title{Orbital obliquities of transiting planets from starspot occultations}
\author{John Southworth$^{1}$, Luigi Mancini$^{2,3}$}

\affiliation{$^{1}$ Astrophysics Group, Keele University, Staffordshire, ST5 5BG, UK \\
             $^{2}$ Max Planck Institute for Astronomy, K\"{o}nigstuhl 17, 69117 -- Heidelberg, Germany \\
             $^{3}$ INAF -- Osservatorio Astrofisico di Torino, via Osservatorio 20, 10025 -- Pino Torinese, Italy}

\shorttitle{Planets transiting starspots}
\shortauthors{Southworth \&  Mancini}

\abs{When a planet passes in front of a starspot during a transit of its host star, it causes a small upward blip in the light curve. Modelling the transit with the starspot allows the size, brightness and position of the spot to be measured. If the same spot can be observed in two different transits, it is possible to track the motion of the spot due to the rotation of the star. The rotation period and velocity of the star (\Prot\ and \Vsini) and the sky-projected orbital obliquity of the system ($\lambda$) can then be determined. If one has three or more observations of the same spot, the {\em true} orbital obliquity ($\psi$) can be measured. We are performing this analysis for a number of cool stars orbited by transiting planets. We present our results so far and compile a catalogue of $\lambda$ and $\psi$ measurements from spot crossing events. The method is particularly useful for cool stars, and is therefore complementary to studies of the Rossiter-McLaughlin effect, which perform better on hotter and faster-rotating stars.}

\begin{document}
\maketitle

\section{Introduction}

The study of extrasolar planets is currently one of the most active areas of research in astrophysics, encompassing observational and theoretical analyses of gas giants, rocky planets, their host stars, and the formation and evolution of planetary systems including our own Solar system. Those planets which transit their host star afford particular excitement as they are the only examples whose masses and radii are measurable, from which additional insight into their formation, structure and dynamical evolution can be gleaned.

A critical problem in planetary astrophysics concerns the genesis of hot Jupiters, which can be defined as giant planets with masses of 0.3\Mjup\ or more, on very tight orbits with orbital periods $\Porb < 10$\,d \citep{Me15aspc}. They are predicted to form beyond the ``ice line'' in a protoplanetary disc \citep[e.g.][]{Lin++96nat,Baruteau+14prpl} so must have jettisoned most of their angular momentum in order to arrive at their current orbits. The two possible causes are smooth inward migration through the protoplanetary disc \citep[e.g.][]{GoldreichTremaine80apj} and gravitational effects such as Kozai-Lidov cycles and planet-planet scattering \citep{FabryckyTremaine07apj,DawsonMurray13apj} combined with tidal evolution \citep{Albrecht+12apj2}. The two migration mechanisms predict different dynamical characteristics for the present-day orbits, in particular orbital obliquity (the relative alignment of the planet's orbital axis and the rotational axis of the host star). Observational constraints on this quantity therefore enable investigations into the formation and evolution mechanisms of giant planets.

The sky-projected orbital obliquity ($\lambda$) can be measured observationally in several ways. Measurement of the Rossiter-McLaughlin effect\footnote{See the TEPCat catalogue \citep{Me11mn} which can be found at {\tt http://www.astro.keele.ac.uk/jkt/tepcat/}} (an anomalous radial velocity of the host star caused by occultation of a portion of its rotating surface) has now been achieved for a total of 90 transiting planets (e.g.\ \citealt{Triaud+10aa,Albrecht+12apj2}) and is effective at moderate stellar rotational velocities. Orbital obliquities can be measured for systems containing faster-rotating stars by the method of Doppler tomography \citep{Cameron+10mn,Cameron+11mn} including in some cases the true orbital obliquity ($\psi$). Both methods fail for slowly-rotating stars (those whose \Vsini\ is not significantly greater than the spectral resolution), which are predominantly cool dwarfs spun down by magnetic braking.

Orbital obliquity measurements for slowly-rotating stars can instead be obtained by tracking the movement of spots on the surfaces of these stars. When occulted by a planet, they imprint an anomalous brightening on the transit light curve whose morphology depends on the spot position, size, and brightness relative to the pristine photosphere of the star. A detection of the positional displacement of one starspot observed during two transits immediately yields a measure of $\lambda$. Furthermore, observation of the position of one spot during three or more different transits allows constraints to be placed on $\psi$ \citep{Nutzman+11apj2,SanchisWinn11apj}, which is not accessible via the Rossiter-McLaughlin effect (but see also \citealt{Cegla+16aa}).

In this work we present a project to expand the number of transiting planetary systems for which orbital obliquity measurements have been achieved, in order to explore the orbital configurations of planets around cool stars. Our work is based on obtaining high-precision photometry of planetary transits through the telescope-defocussing method \citep{Alonso+08aa,Me+09mn2,Me+09apj,Me+09mn}, and on modelling the resulting light curves using the {\sc prism+gemc} code \citep{Tregloan++13mn,Tregloan+15mn}. Below we review our work so far and assemble a catalogue of obliquity measurements obtained using the spot-tracking method.

\section{WASP-19}

\begin{figure}[t]
\centering
\includegraphics[width=\linewidth]{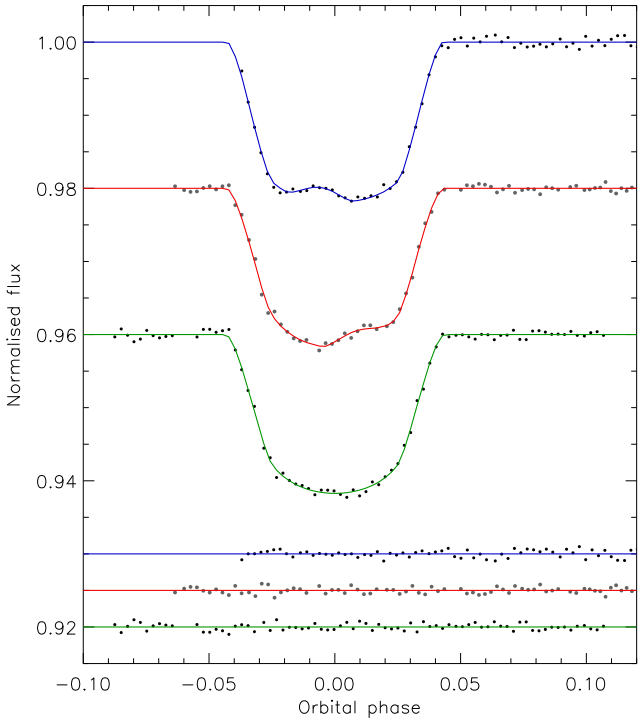}
\caption{Light curves of WASP-19 (filled circles) compared to the best-fitting {\sc prism+gemc} models (solid lines). The residuals are positioned at the base of the figure, offset from zero.}
\label{fig:w19}
\end{figure}

The WASP-19 system \citep{Hebb+10apj} contains a giant planet with a very short orbital period (0.788\,d), a mass of $1.139 \pm 0.036$\Mjup\ and a radius of $1.410 \pm 0.021$\Rjup\ \citep{Mancini+13mn2}, orbiting a star moderately cooler than the Sun ($\Teff = 5460 \pm 90$\,K). We observed three transits over a period of five nights (Fig.\,\ref{fig:w19}) using the ESO New Technology Telescope (NTT), serendipitously detecting the occultation of the same spot on two successive nights \citep{Tregloan++13mn}. We introduced the {\sc prism} code, which uses the pixellation method and a Cartesian co-ordinate system to generate a synthetic light curve of a spotted star being transited by a planet, and the {\sc gemc} code, which fits this model to observational data using a differential-evolution MCMC approach \citep{Terbraak06}. This approach allowed measurement of the star's rotation period ($\Prot = 11\fd76 \pm 0\fd09$ at a latitude of 35\degr) and of $\lambda = 1\fo0 \pm 1\fo2$. The last value is a significant improvement on that obtained from spectroscopic observations of the Rossiter-McLaughlin effect ($\lambda = 4\fo6 \pm 5\fo2$; \citealt{Hellier+11apj}). The NTT turns out to be a very good telescope for obtaining high-precision photometry for planet hosts with a good comparison star within 4\,arcmin \citep{TregloanMe13mn}.

\section{Qatar-2}

\begin{figure}[t]
\centering
\includegraphics[width=\linewidth]{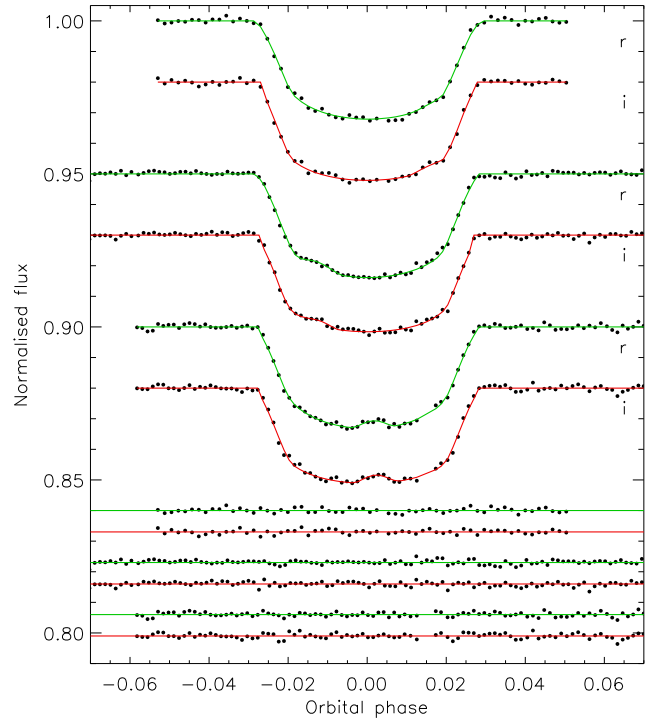}
\caption{The GROND $r$- and $i$-band light curves of Qatar-2 (filled circles) compared to the best-fitting {\sc prism+gemc} models (solid lines). The residuals are positioned at the base of the figure, offset from zero.}
\label{fig:q2}
\end{figure}

The star Qatar-2\,A is a cool dwarf ($\Teff = 4645 \pm 50$\,K) found by \citet{Bryan+12apj} to host a planet, Qatar-2\,b, orbiting every 1.34\,d. The planet has a high density, with a mass of $2.494 \pm 0.054$\Mjup\ and a radius of $1.254 \pm 0.013$\Rjup\ \citep{Mancini+14mn}. We observed three transits using the GROND imager on the 2.2\,m telescope at ESO La Silla, which is capable of observing in four optical bands simultaneously (similar to the SDSS $griz$ bands), plus three transits from other telescopes \citep{Mancini+14mn}. All transit light curves exhibited starspot anomalies. Two of the transits observed with GROND showed the same starspot (Fig.\,\ref{fig:q2}), from which we calculated $\Prot = 14\fd8 \pm 0\fd3$ at a latitude of 16\degr, and $\lambda = 4\fo3 \pm 4\fo5$. This is the only orbital obliquity measurement for Qatar-2 currently published, but spectroscopic observations of the Rossiter-McLaughlin effect confirm this result (Esposito et al., in prep) albeit with larger errorbars.

\section{WASP-6}

\begin{figure}[t]
\centering
\includegraphics[width=\linewidth]{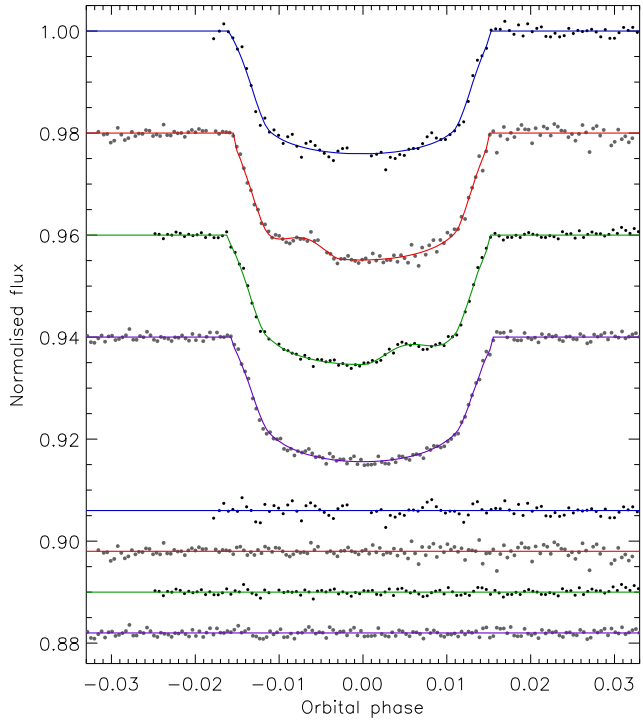}
\caption{Light curves of WASP-6 (filled circles) compared to the best-fitting {\sc prism+gemc} models (solid lines). The residuals are positioned at the base of the figure, offset from zero.}
\label{fig:w6}
\end{figure}

WASP-6 comprises a low-density hot Jupiter (mass $0.485 \pm 0.028$\Mjup\ and radius $1.230 \pm 0.037$\Rjup) orbiting a cool star ($\Teff = 5383 \pm 41$\,K) every 3.36\,d \citep{Gillon+09aa3}. It is the object which sparked our project, after two transits observed 27\,d apart in 2009 showed spot crossing events. A total of four transits were observed using the Danish 1.54\,m Telescope at ESO La Silla \citep{Tregloan+15mn} of which the middle two showed the signature of occulted starspots (Fig.\,\ref{fig:w6}). Under the assumption that they were due to the same spot, which had travelled slightly more than once round the star in the interval between these two transits, we found $\Prot = 23\fd80 \pm 0\fd15$ at a latitude of 14\degr\ and $\lambda = 7\fo2 \pm 3\fo7$. These values are consistent with -- and a significant improvement on -- spectroscopic observations of the Rossiter-McLaughlin effect, which indicate $\lambda = \er{14}{14}{18}$\,degrees \citep{Gillon+09aa3}.

\section{WASP-41}

\begin{figure}[t]
\centering
\includegraphics[width=\linewidth]{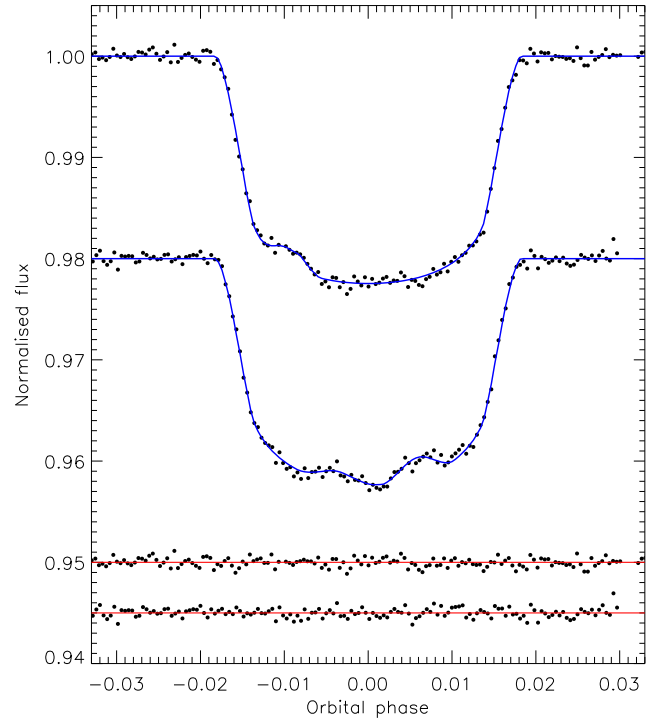}
\caption{Light curves of WASP-41 (filled circles) compared to the best-fitting {\sc prism+gemc} models (solid lines). The residuals are positioned at the base of the figure, offset from zero.}
\label{fig:w41}
\end{figure}

The WASP-41 system shows deep transits of an active G8\,V star \citep{Maxted+11pasp} with $\Teff = 5546 \pm 33$\,K \citep{Mortier+13aa}, caused by a planet of mass $0.977 \pm 0.026$\Mjup, radius $1.178 \pm 0.018$\Rjup\ and orbital period 3.05\,d \citep{Me+16mn}. \citet{Neveu+16aa} have found a second planet with a much longer period of $421 \pm 2$\,d which is not known to transit but was identified from radial-velocity monitoring of the host star. We observed four transits by the short-period planet using the Danish Telescope, of which two consecutive transits showed anomalies due to the same starspot (Fig.\,\ref{fig:w41}) and allowed us to determine $\Prot = 18\fd6 \pm 1\fd5$ at a latitude of 20\degr, and $\lambda = 6\degr \pm 11\degr$. This obliquity measurement improves on the value from the Rossiter-McLaughlin effect of $\lambda = \er{29}{10}{14}$\,degrees \citep{Neveu+16aa}.

\section{WASP-52}

\begin{figure}[t]
\centering
\includegraphics[width=\linewidth]{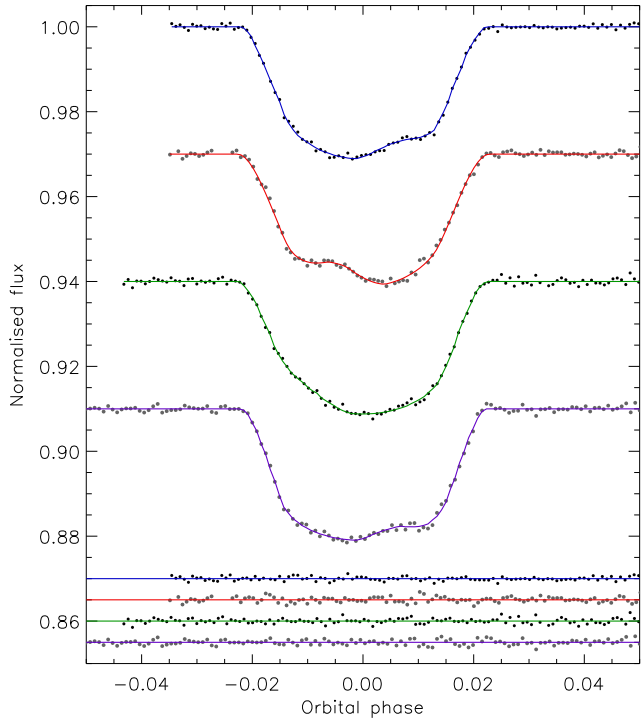}
\caption{Light curves of WASP-52 (filled circles) compared to the best-fitting {\sc prism+gemc} models (solid lines). The residuals are positioned at the base of each panel, offset from zero.}
\label{fig:w52}
\end{figure}

WASP-52 comprises a low-density planet (mass $0.46 \pm 0.02$\Mjup, radius $1.27 \pm 0.03$\Rjup) on a 1.75\,d period orbit around a cool dwarf ($\Teff = 5000 \pm 100$\,K; \citealt{Hebrard+13aa}). We observed eight transits in this system, using the Danish Telescope, GROND, the 1.23\,m telescope at Calar Alto Observatory, Spain, and the 1.5\,m Cassini telescope at the Observatory of Bologna, Italy \citep{Mancini+16}. Five of the transits show starspot anomalies (Fig.\,\ref{fig:w52}). Four of these appear to be due to the same spot, and are each separated by 14--15\,d. We find a sky-projected orbital obliquity of $\lambda = 3\fo8 \pm 8\fo4$, as well as $\Prot = 13\fd1 \pm 0\fd4$ at a latitude of 18\degr. Having observations of the same spot at four different positions on the rotating stellar surface has allowed us to determine the {\em true} orbital obliquity of the system to be $\psi = 20\degr \pm 50\degr$. Its precision is limited by the uncertainties in the spot latitudes, which show degeneracies for partially-eclipsed spots as these could be located either above or below the transit cord. To our knowledge this is the first measurement of $\psi$ purely from spot crossings.

\section{Summary}

The true and sky-projected orbital obliquities of planetary systems retain information on their formation and dynamical evolution histories, albeit eroded by tidal effects in these short-period binary systems. Measurements of $\lambda$ are routinely achieved using spectroscopic observations of the Rossiter-McLaughlin effect for moderately-rotating host stars. Fast rotation compromises this process by making the radial velocity measurements uncertain, but such systems can be analysed using Doppler tomography to obtain $\lambda$ and in some cases $\psi$. However, host stars showing no detectable rotational broadening in their spectra are observationally inaccessible by these methods \citep[e.g.][]{Albrecht+11apj2}. As these are preferentially cool stars, due to the effect of magntic braking, there exists a gap in our knowledge of the orbital configurations of transiting planet host stars. This gap can be filled by tracking the motion of spots as the stars rotate, using the transiting planets as moving masks to resolve the stellar surfaces.

\begin{figure*}
\includegraphics[width=\linewidth,angle=0]{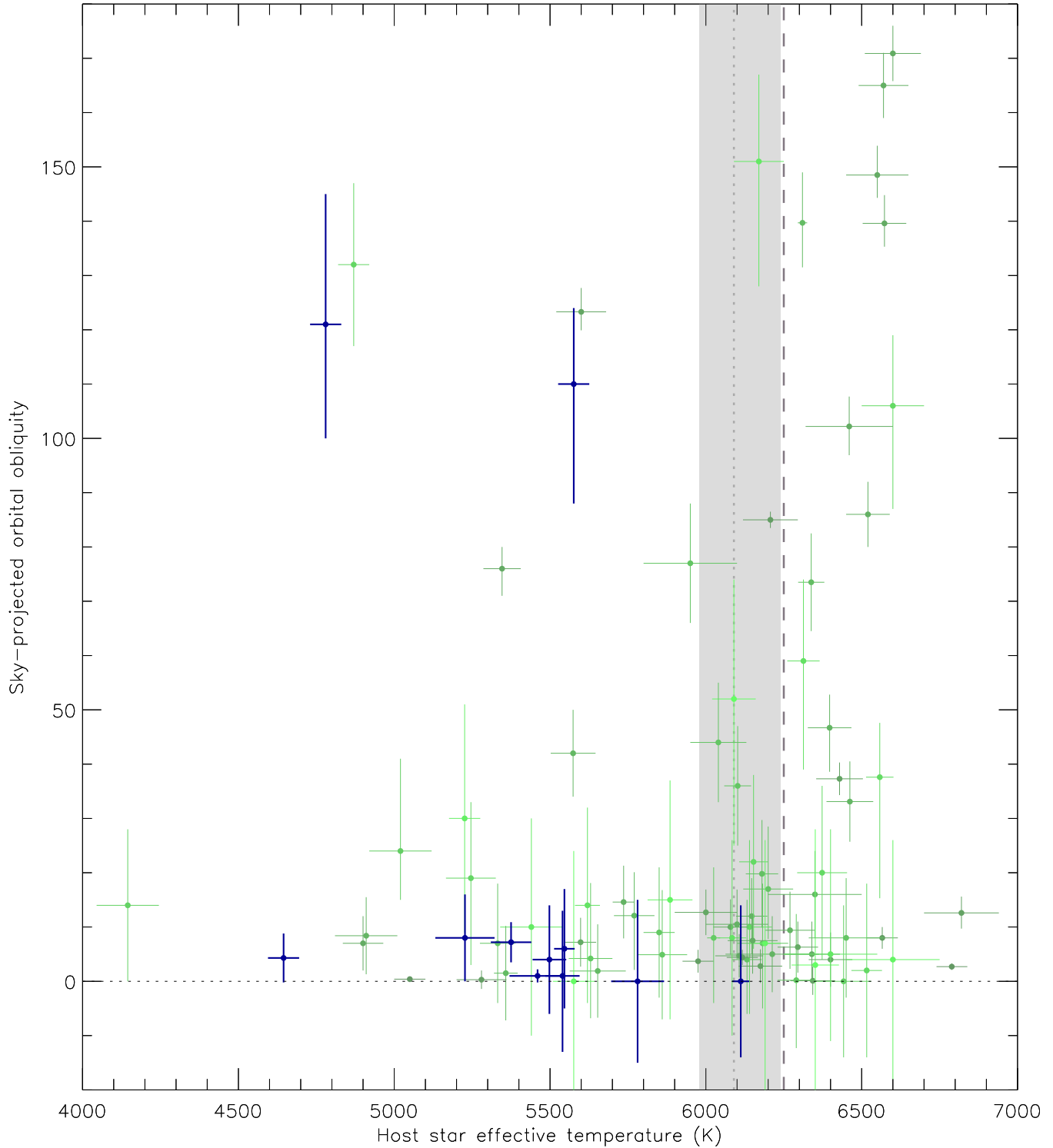}
\caption{Plot of sky-projected orbital obliquity, $\lambda$, versus the \Teff\ of the host star. Green lines show measurements from the Rossiter-McLaughlin effect, and blue lines show those from spot-crossing analyses.
The data for this figure were taken from TEPCat \citep{Me11mn} on 2016/07/27. The \Teff\ values proposed as the boundaries between stars with strong and weak tidal coupling to the planet are shown using a grey dashed line \citep[6250\,K;][]{Winn+10apj3} and a grey shaded region with a dotted line \citep[\er{6090}{150}{110}\,K;][]{Dawson14apj}. Published values of $\lambda$ have been adjusted by $\pm$180$^\circ$ to bring them into the interval [$0^\circ$,$180^\circ$] following the discussion by \citet{CridaBatygin14aa}.}
\label{fig:tl}
\end{figure*}

This work presents a project to obtain transit light curves of cool and magnetically active planet host stars, from which the instantaneous positions of starspots can be measured and thus the movement of these spots tracked. This process can yield both $\lambda$ and $\psi$, depending on the number of transits observed, using essentially geometric arguments. We have measured $\lambda$ for five planetary systems (WASP-19, Qatar-2, WASP-4, WASP-41, WASP-52) and also presented the first measurement of $\psi$ using this method (for WASP-52). These results are put into context in Fig.\,\ref{fig:tl} and the full sample of $\lambda$ and $\psi$ measurements from spot-crossing analyses are collected in Table\,\ref{tab:lp}. This catalogue is maintained as part of the TEPCat catalogue of transiting planets \citep{Me11mn} which can be found at {\footnotesize{\tt http://www.astro.keele.ac.uk/jkt/tepcat/}}. Future work will include increasing the number of entries in Table\,\ref{tab:lp} and in exploiting the longitudinal separation of European and Chilean observatories for observing multiple consecutive transits of planets orbiting magnetically-active stars. As part of this work we will refine the measured physical properties of the systems using the {\it Homogeneous Studies} methodology \citep{Me08mn,Me09mn,Me10mn,Me11mn,Me12mn}.

\begin{table*}[t]
\centering
\caption{Published measurements of the sky-projected and true orbital obliquities of transiting planetary systems from spot-tracking analyses.}
\begin{tabular}{l c c c l}
\hline
System    & Host star \Teff\ (K) & $\lambda$ ($^\circ$)         & $\psi$ ($^\circ$)               & Reference                 \\
\hline
HAT-P-11  & $4780 \pm  50$ &\er{105}{16}{12} or \er{121}{24}{21}&\er{106}{15}{11} or \er{97}{8}{4}& \citet{SanchisWinn11apj}  \\
HATS-2    & $5227 \pm  95$ & $8 \pm 8$                          &                                 & \citet{Mohler+13aa}       \\
Kepler-17 & $5781 \pm  85$ & $0 \pm 15$                         &                                 & \citet{Desert+11apjs}     \\
Kepler-30 & $5498 \pm  54$ & $-1 \pm 10$ or $4 \pm 10$          &                                 & \citet{Sanchis+12nat}     \\
Kepler-63 & $5576 \pm  50$ & \er{-110}{22}{14}                  & \er{145}{9}{14}                 & \citet{Sanchis+13apj2}    \\
Qatar-2   & $4645 \pm  50$ & $4.3 \pm 4.5$                      &                                 & \citet{Mancini+14mn}      \\
WASP-4    & $5540 \pm  55$ & \er{-1}{14}{12}                    &                                 & \citet{Sanchis+11apj}     \\
WASP-6    & $5375 \pm  65$ & $7.2 \pm 3.7$                      &                                 & \citet{Tregloan+15mn}     \\
WASP-19   & $5460 \pm  90$ & $1.0 \pm 1.2$                      &                                 & \citet{Tregloan++13mn}    \\
WASP-41   & $5546 \pm  33$ & $6 \pm 11$                         &                                 & \citet{Me+16mn}           \\
WASP-52   & $5020 \pm 100$ & $3.8 \pm 8.4$                      & $20 \pm 50$                     & \citet{Mancini+16}        \\
WASP-85   & $6112 \pm  27$ & $0 \pm 14$                         &                                 & \citet{Mocnik+16aj}       \\
\hline
\end{tabular}
\label{tab:lp}
\end{table*}

\section*{Acknowledgments}

JS acknowledges financial support from the Leverhulme Trust in the form of a Philip Leverhulme Prize. We thank Dr.\ Jeremy Tregloan-Reed for supplying the light curves and models used to generate Figs.\ \ref{fig:w19} and \ref{fig:w6}.

\bibliographystyle{cs19proc}

\end{document}